\documentclass{jfm}
\usepackage{natbib}
\usepackage{graphicx}
\usepackage{amsbsy} 
\usepackage{amsmath}
\usepackage{amssymb}
\usepackage{amsfonts}
\usepackage{psfrag}
\usepackage{wrapfig}
\usepackage{textcomp}
\usepackage{bm}
\usepackage[usenames]{color}

\newcommand\Pra{\mbox{\textrm{Pr}}} % Prandtl number, cf TeX's \Pr product
\newcommand\Ra{\mbox{\textrm{Ra}}} % Rayleigh number
\newcommand\Nu{\mbox{\textrm{Nu}}} % Nusselt number 

\def\gtwid{\mathrel{\raise.3ex\hbox{$>$\kern-.75em\lower1ex\hbox{$\sim$}}}}
\def\alt{\mathrel{\raise.3ex\hbox{$<$\kern-.75em\lower1ex\hbox{$\sim$}}}}

\def\agt{\mathrel{\raise.3ex\hbox{$>$\kern-.75em\lower1ex\hbox{$\sim$}}}}

\newcommand{\be}{\begin{equation}}
\newcommand{\ee}{\end{equation}}

\begin{document}

\title[Logarithmic temperature profiles of turbulent Rayleigh-B\'enard convection]{Logarithmic temperature profiles of turbulent Rayleigh-B\'enard convection in the classical and ultimate state for a Prandtl number of 0.8}

\author{Guenter Ahlers$^{1,2,5}$, Eberhard Bodenschatz$^{2,3,4,5}$, and Xiaozhou He$^{2,5}$}

\affiliation{
$^{1}$Department of Physics, University of California, Santa Barbara, CA 93106, USA
\\
$^{2}$Max Planck Institute for Dynamics and Self-Organization (MPIDS), 37077 G\"ottingen, Germany
\\
$^{3}$Institute for Nonlinear Dynamics, University of G\"ottingen, 37077 G\"ottingen, Germany
\\
$^{4}$Laboratory of Atomic and Solid-State Physics and Sibley School of Mechanical and Aerospace Engineering, Cornell University, Ithaca, New York 14853, USA
\\
$^{5}$International Collaboration for Turbulence Research
}

\date{\today}

\maketitle

\vskip 0.2in

\begin{abstract}
We report on experimental determinations of the temperature field in the interior (bulk) of turbulent Rayleigh-B\'enard convection (RBC) for a cylindrical sample with an aspect ratio (diameter $D$ over height $L$) equal to 0.50, both in the classical and in the ultimate state. The measurements are for the \Ra\ range from $6\times 10^{11}$ to $10^{13}$ in the classical and  for $\Ra = 7\times 10^{14}$ to $1.1\times 10^{15}$ (our maximum accessible \Ra) in the ultimate state. The Prandtl number was åclose to 0.8. Although to lowest order the bulk is often presumed to be isothermal in the time average, we find a ``logarithmic layer" (as already reported by \cite{ABFGHLSV12}) in which the reduced temperature $\Theta = [\langle T(z) \rangle - T_m]/\Delta T$ (with $T_m$ the mean temperature and $\Delta T$ the applied temperature difference) varies as $A*ln(z/L) + B$ or $A^\prime *ln(1-z/L) + B^\prime$ with the distance $z$ from the bottom plate of the sample. In the classical state the amplitudes $-A$ and $A^\prime$ are equal within our resolution, while in the ultimate state there is a small difference with $-A/A^\prime \simeq 0.95$. For the classical state the width of the log layer is about $0.1L$, the same near the top and the bottom plate as expected for a system with reflection symmetry about its horizontal mid plane. For the ultimate state the log-layer width is larger, extending through most of the sample, and slightly asymmetric about the mid plane. Both amplitudes $A$ and $A^\prime$ vary with radial position $r$, and this variation can be described well by $A = A_0 [(R - r)/R]^{-0.65}$ where $R$ is the radius of the sample. In the classical state these results are in good agreement with direct numerical simulations (DNS) for $\Ra = 2\times 10^{12}$; in the ultimate state there are as yet no DNS.  The amplitudes $-A$ and $A^\prime$ varied as $\Ra^{-\eta}$, with $\eta \simeq 0.12$ in the classical and  $\eta \simeq 0.18$ in the ultimate state. A close analogy between the temperature field in the classical state and the ``Law of the Wall" for the time averaged down-stream velocity in shear flow is discussed. Although there are strong similarities, the \Ra\ dependence of $A$ differs from the universal (Reynolds-number independent) von K\'arm\'an constant in shear flow. In the ultimate state the result $\eta \simeq 0.18$ differs from the prediction $\eta \simeq 0.043$ by \cite{GL12}. 

\end{abstract}

\section{Introduction}
\label{sec:intro}

Turbulent convection in a fluid contained between two parallel horizontal plates and heated form below (Rayleigh-B\'enard convection or RBC) has ben studied intensely for many decades both theoretically and experimentally [for various reviews, see \cite{Ka01,Ah09, AGL09, LX10,CS12}]. For this system most of the vertical temperature change $\Delta T \equiv T_b - T_t$ between the bottom temperature $T_b$ and the top temperature $T_t$ occurs over two thin thermal boundary layers, one adjacent to each of the top and the bottom plate. For typical cases where the adiabatic gradient is negligible, the bulk of the sample was long considered to have a constant  temperature $T_m \equiv (T_b + T_t)/2$ in the time average, although the local temperature was understood to fluctuate vigorously. This model, which is due to \cite{Ma54}  and \cite{Pr54,Pr59} [see also the review by \cite{Sp71}, Sec. 5], has served well for several decades to explain many features of this system at a semi-quantitative level.

More detailed measurements during the last two decades revealed  that, while the temperature drop across the thermal boundary layers accounted for the major part of $\Delta T$ \cite[]{LX98}, there were significant measurable temperature variations in the bulk of the sample [see, for instance, \cite{TBL93,BA07_EPL}]. In the absence of more detailed information, these temperature profiles often were assumed to correspond to a constant temperature gradient, with the total temperature drop across the bulk only a few percent of $\Delta T$ and dependent on the Rayleigh number, the Prandtl number, and the radial location within the sample. These bulk gradients usually were attributed to thermal dissipation due to plumes which emanate from the top and bottom boundary layers. The plumes are carried by, and due to their buoyancy in turn drive, a large-scale circulation (LSC)  so that they rise or fall through the bulk; but we are not aware of a detailed theoretical explanation, based on plumes, of the bulk temperature field.

Recently it was discovered experimentally \cite[]{ABFGHLSV12} and confirmed by direct numerical simulation (DNS) [\cite{SLV11}, as reported by \cite{ABFGHLSV12}]  that the bulk temperature profiles in a cylindrical sample of diameter $D$ and aspect ratio $\Gamma \equiv D/L = 0.5$ are far richer than had been anticipated. It turns out that the time averaged bulk temperature  
\be
\Theta(z,r) \equiv [\langle T(z,r,t)\rangle - T_m]/\Delta T
\label{eq:Theta}
\ee
 (we denote the average over the time $t$ by $\langle ... \rangle$) varies logarithmically with the distance $z$ from the bottom or $L-z$ from the top plate over a significant fraction of the sample height $L$, {\it i.e.}
\be
\Theta = A(r)*ln(z/L) + B(r)
\label{eq:Theta2}
\ee      
near the bottom and
\be
\Theta = A^\prime(r)*ln(1 - z/L) + B^\prime(r)
\label{eq:Theta3}
\ee      
near the top plate (here $z$ is the distance from the bottom plate and $r$ is the radial distance from the vertical center line). It was found from DNS that the amplitudes $A$ and $A^\prime$ decrease as the radial location is varied from near the side wall toward the sample interior. Although in the RBC system typical values of $|\Theta|$ are less that 0.05 because most of the temperature drop is, as mentioned, across thermal boundary layers, the very existence of logarithmic profiles provokes a comparison with the ``Law of the Wall" for the time-averaged down-stream (``stream-wise") velocity in turbulent shear flows first enunciated by \cite{Pr25,Pr32} and \cite{Ka30} and studied intensely ever since [for recent reviews, see for instance \cite{Po00,MMMNSS10,SMM11}].

Before discussing the results of the present work, we note that turbulent RBC can exist in two or more distinct states, as first discussed by \cite{Kr62} and \cite{Sp71} and more recently by \cite{GL11}. For \Ra\ values below a characteristic $\Ra^*$ it was expected that the system has viscous boundary layers (BLs) adjacent to the top and bottom plates, as well as the side walls, established by the LSC or, in the absence of an LSC, by velocity fluctuations on length scales smaller than $D$. These viscous BLs then are believed to co-exist with the above-mentioned thermal BLs adjacent to the plates. We refer to this state as the ``classical" state since it is the one that has been studied experimentally for many decades; it is the state envisioned by early workers such as \cite{Ma54} and \cite{Pr54,Pr59}. For larger \Ra\ it was expected that the viscous BLs become turbulent due to the shear applied to them by the LSC or the fluctuations. This state is of particular interest because it is expected to be asymptotic in the sense that it will prevail up to arbitrarily large \Ra; thus it is referred to as the ``ultimate state" \cite[]{CCCHCC97} and its properties, once studied above but near $\Ra^*$, can in principle be extrapolated to the very high Rayleigh numbers relevant to various astrophysical and geophysical phenomena that are not attainable in the laboratory. 

Although \cite{Kr62} had predicted the as yet unattainable values $\Ra^* = {\cal O}(10^{21})$ or larger, a more realistic estimate \cite[]{GL02} based on more recent bulk Reynolds-number measurements by  \cite{QT01a} and the RBC model of \cite{GL00} gave $\Ra^* \simeq 10^{14}$ for a Prandtl number near one. Experimentally it was found recently for a cylindrical sample with $\Gamma = 0.50$ that the transition to the ultimate state takes place gradually over a range of \Ra, from $\Ra_1^* \simeq 10^{13}$ to $\Ra_2^* \simeq 5\times10^{14}$ \cite[]{HFNBA12,AHFB12,HFBA12}. Earlier heat-transport measurements \cite[]{CCCHCC97,CCCCH01,RGKS10} had suggested that this transition takes place already near $\Ra \simeq 10^{11}$; but in our view such a low value of $\Ra^*$ would be inconsistent with the shear-induced boundary-layer transition to turbulence associated with the ultimate-state transition because the shear Reynolds number is too small. There is, however, at present no alternative explanation of the transition that is clearly evident in these early heat-transport measurements but absent in other similar measurements \cite[]{Wu91,NSSD00} [for a recent detailed re-examination of all of those data, see \cite{AHFB12}].

From the viewpoint of the present paper the ultimate state is of particular interest because \cite{GL11} predicted for the Boussinesq system that each of the  two turbulent kinetic BLs extends all the way to the sample mid plane at $z = L/2$, thus essentially leaving no ``bulk". In a subsequent paper \cite[]{GL12} they showed that these turbulent BLs produce logarithmic temperature profiles. 
This implies that, at least in the ideal system, there should be logarithmic temperature profiles extending away from both the top and the bottom plate, and that these profiles should meet at the horizontal mid plane of the sample. This interesting prediction was the original motivation for the present work. Thus, we were prepared to find the Grossmann-Lohse log profiles in the ultimate state for $\Ra > \Ra^*_2$. However, to our great surprise  we found very similar profiles also in the classical state for $\Ra < \Ra^*_1$. 

We assume that the logarithmic temperature distributions in classical RBC have an origin not unlike the one responsible for the logarithmic profiles of the time-averaged down-stream velocity in shear-flow systems such as plane Couette, Couette-Taylor, or pipe flow. However, in the RBC case the profile is not dependent upon the presence of shear along the walls or plates; rather it seems likely to us that it is associated with the presence and vertical distribution of thermal excitations which are usually referred to as plumes. This is demonstrated convincingly in a recent numerical study \cite[]{PMGL13} of a two-dimensional RBC system of large aspect ratio and with periodic lateral boundary conditions where there is no sidewall shear. The authors found that there are several LSC cells, each roughly with a diameter equal to the height. While there were logarithmic  vertical  temperature profiles at all horizontal locations, the amplitude $A$ had a pronounced maximum at the up-flow locations. It is also at these locations that the concentration of rising plumes emitted from the thermal BLs is largest, thus showing a correlation between plume density and log-profile amplitude. Thus it seems likely that the plumes in RBC play a role similar to that of the velocity excitations (``coherent eddies") in shear flows. However, to our knowledge a detailed theory for RBC leading from plumes to logarithmic temperature profiles is still missing. 

In the present paper we present more extensive experimental results for the logarithmic profiles of the temperature in the interior of turbulent convection in a slightly modified version (HPCF-IIg) of the cylindrical sample of aspect ratio 0.50 used before (HPCF-IIf) 
\cite[]{ABFGHLSV12}. Both the axial and the radial dependences of the coefficients of Eqs.~\ref{eq:Theta2} and \ref{eq:Theta3} were studied in the classical state over the \Ra\ range from $6\times 10^{11}$ to $10^{13}$ and in the ultimate state from $\Ra = 7\times 10^{14}$ to $1.1\times 10^{15}$ (our maximum accessible \Ra). 

After defining the parameters relevant to this work in Sec.~\ref{sec:paras}, we give a brief description of the apparatus in Sec.~\ref{sec:app}. A detailed discussion of the temperature probes (thermistors) used in this work is provided in Sec.~\ref{sec:thermistors}. There the details of the locations of these sensors are given.
The results for the classical state are presented in Sec.~\ref{sec:classical}. At the end of that section, in subsection \ref{sec:compLoW},  we discuss  the analogies, similarities, and differences between RBC and shear flows. In Sec.~\ref{sec:ult} we present results for the ultimate state. 
We conclude this paper with a brief summary in Sec.~\ref{sec:summary}.   

\section{Relevant parameters}
\label{sec:paras}

For a given sample geometry, the state of the system depends on two dimensionless variables.
The first is the Rayleigh number $\Ra$, a dimensionless form of the temperature difference $\Delta T = T_b-T_t$ between the bottom ($T_b$) and the top ($T_t$) plate. It is given by
\begin{equation}
\Ra=\frac{g\alpha\Delta T L^3}{\kappa \nu}\mbox{.}
\label{eq:Ra}
\end{equation}
Here, $g$, $\alpha$, $\kappa$ and $\nu$ denote the gravitational acceleration, the isobaric thermal expansion coefficient, the thermal diffusivity, and the kinematic viscosity respectively.  The second is the Prandtl number
\be
\Pra=\nu/\kappa\ .
\ee
Unless stated otherwise, all fluid properties are evaluated at the mean temperature $T_m = (T_b + T_t)/2$.

The vertical heat transport from the bottom to the top plate is expressed in dimensionless form by the Nusselt number
\begin{equation}
Nu=\frac{\lambda_{eff}}{\lambda}
\label{eq:Nu}
\end{equation}
where the effective conductivity
\begin{equation}\label{eq:lambda_eff}
\lambda_{eff}=Q L/(A \Delta T)
\end{equation}
and where $Q$ is the heat flux and $A$ is the cross sectional area of the cell.

For samples in the shape of right-circular cylinders like those used here, a further parameter defining the geometry is needed and is the aspect ratio $\Gamma \equiv D/L$ where $D$ is the sample diameter.

In the axial direction we shall use $z/L$ in the lower half and $1 - z/L$ in the upper half of the sample as the relevant dimensionless coordinate. For the radial position we shall use the parameter 
\be
\xi \equiv (R-r)/R
\ee
where $r$ is the radial distance from the center line in cm and $R= 56.0$ cm is the radius of the sample.

\section{Apparatus and Procedures}
\label{sec:app+proc}

\subsection{Apparatus}
\label{sec:app}

The apparatus was described in detail elsewhere \cite[]{AFB09,AHFB12}. In the present work we used version HPCF-IIg, a slight modification of version HPCF-IIf listed in Table 1 of \cite{AHFB12}. As will become evident below, the two versions differed only in the number of thermistors that had been installed to probe the temperature profiles in the sample interior. 
Here we describe only  the main features of the overall facility,  and then, in Sec.~\ref{sec:thermistors},  give details about  thermistors that were added for the present work.

\begin{figure}
\centerline{\includegraphics[width=0.65 \textwidth]{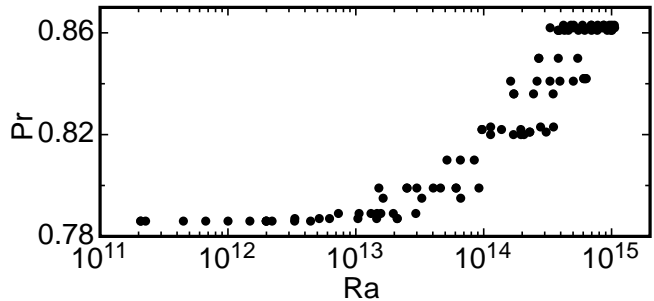}}
\caption{Typical values of \Pra\ as a function of \Ra. The discrete levels of \Pra\ are associated with corresponding discrete pressure values.}
\label{fig:Pr}
\end{figure}

The sample cell (known as the High Pressure Convection Facility or HPCF) was located inside a pressure vessel known as the ``Uboot of G\"ottingen" which could be pressurized with various gases. For the present work we used sulfur hexafluoride (SF$_6$) at an average temperature $T_{m} = 21.5$\textcelsius\ and at pressures that ranged from 2.0 to 
17.7 bars, resulting in a Prandtl number that varied over the relatively narrow range from 0.78 to 0.86 over the \Ra\ range from $10^{11}$ to $10^{15}$ as shown in Fig.~\ref{fig:Pr}.  The Uboot had a volume of about 25 m$^3$, and about 2000 kg of SF$_6$ were required to fill it to the maximum pressure. The sample cell had  copper plates at the top and bottom that were separated by a distance $L = 224$ cm.
In the lateral direction it had a cylindrical side wall made of a Plexiglas tube of inner diameter $D=112$ cm and wall thickness 0.95 cm, yielding an aspect ratio  $\Gamma = 0.500$. The cell was completely sealed; it had a small-diameter tube entering it through the side wall at mid height to permit the fluid to enter from the Uboot. During measurements this tube was sealed by a remotely controlled valve.  
Various thermal shields outside the sample cell prevented parasitic heat losses or inputs. 

During all measurements, the temperatures at the top and bottom plates were held constant by digital feedback loops. One such loop controlled the power going to the bottom-plate heater. The top-plate temperature was held constant by a temperature-controlled cooling-water circuit. For each set of conditions (pressure, $T_m$, and $\Delta T$), measurements of all thermometers and heater currents were made at time intervals of about 12 sec, and over a period of typically one day. Typically  the data taken during the first eight hours were discarded to avoid the influence of transients, and the remainder was analyzed.

\subsection{Thermistors used in the sample interior}
\label{sec:thermistors}

\subsubsection{Thermistor locations}
\label{sec:types}

\begin{figure}
\centerline{\includegraphics[width=0.65 \textwidth]{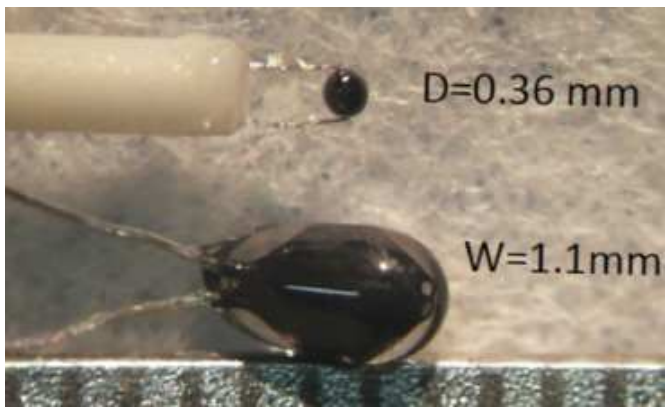}}
\caption{Photograph of the thermistors used in this work. The distance between the lines on the scale is 0.5 mm.}
\label{fig:pictures}
\end{figure}

We used the two types of thermistors illustrated in Fig.~\ref{fig:pictures}. They were inserted to various depths into the sample interior. They differed in size and thus in thermal response time. While the time response is important for fluctuation measurements, for the present work of time-averaged temperature measurements both served equally well and the two types gave equivalent results.

The larger thermistors, to be referred to as T$_1$,  were Honeywell model  112-104KAJ-B01. They were glass encapsulated, and the glass capsule had an outside minor diameter of 1.1 mm and a length of about 1.8 mm. Their Platinum alloy leads had a diameter of 0.10 mm. They were strong enough to be passed directly through a hole in the plexiglas side wall of the convection cell, and to suspend the thermistor at a distance of either 0.1 or 1.0 cm from the inside of the side wall. The hole in the wall was then sealed with silicone glue.

The smaller thermistors were  Honeywell model 111-104HAK-H01, to be called T$_2$. They had  diameters of 0.36 mm and also were glass encapsulated. Thermistors T$_2$ had Platinum-alloy leads of 0.02 mm diameter. The leads were passed, one each, through two 0.13 mm diameter holes parallel to the axis of 0.8 mm diameter ceramic rods (Omega ceramic thermocouple insulators type TRA-005132). The rods were placed in holes of 0.9 mm diameter in the plexiglas side wall of the convection chamber and sealed to the side wall with silicone glue. 

\begin{table}
\begin{center}
\begin{tabular}{cccccccc}
\hline
column  & $d$ ($\mu$m) & $R - r (cm)$ & $\xi \equiv$ & $\theta$ (rad) & N$_{th}$ & HPCF  &Thermistor \\ 
 & & &  (R - r)/R  & & & version & version \\
\hline
$V_0$ & 1140 & 1.0  & 0.0180 & 1.505 & 8 & IIf+g & T$_1$ \\
$V_0^\prime$ & 360 & 1.0  & 0.0180 & 1.505 & 4 & IIg & T$_2$ \\
$V_1$ & 1140 & 1.0  & 0.0180 & 4.646 & 8 & IIf+g & T$_1$ \\
$V_1^\prime$ & 360 & 1.0  & 0.0180 & 4.646 & 4 & IIg & T$_2$ \\
$V_2$ & 360 & 2.0  & 0.0356 & 1.095 & 5 & IIg & T$_2$ \\
$V_3$ & 360 & 2.0  & 0.0356 & 4.236 & 5 & IIg & T$_2$ \\
$V_4$ & 360 & 4.0  & 0.0716 & 0.963 & 5 & IIg & T$_2$ \\
$V_5$ & 360 & 4.0  & 0.0716 & 4.105 & 5 & IIg & T$_2$ \\
$V_6$ & 360 & 15.0 & 0.2680 & 0.518 & 5 & IIg & T$_2$ \\
$V_7$ & 360 & 15.0 & 0.2680 & 3.660  & 5 & IIg & T$_2$ \\
\end{tabular}
\end{center}
\caption{The radial and azimuthal locations of the thermistor columns. The thermistor diameter is given in the table column labeled $d$. The angle $\theta$  is the azimuthal location of the thermistor column relative to an arbitrary origin. Version HPCF-IIf was used by \cite{ABFGHLSV12} and was tilted slightly, with its axis at an
angle of 14 mrad relative to gravity. Version HPCF-IIg was used for the present work and was leveled relative to gravity to within
$10^{-4}$ rad. The table column labeled $N_{th}$ gives the number of thermistors in each thermistor column.}
\label{tab:location}
\end{table}

\begin{table}
\begin{center}
\begin{tabular}{ccccc}
\hline
$z$ (cm)  & $z/L$ & $V_{0}$, $V_{1}$ & $V_{0}^\prime$, $V_{1}^\prime$ & All others \\ 
\hline
4.0 &  0.0178   & X  &     & X \\
6.1 & 0.0272   & X  &     &   \\
8.1 & 0.0362   & X  &     & X \\
12.1 & 0.0540 & X  &     &   \\
16.1 & 0.0719 & X  &     & X \\
32.2 & 0.1438 & X  &     & X \\
64.2 & 0.2866 & X  &     &   \\
110.5 & 0.4933 & X&     & X \\
188.8 & 0.8428 &   & X  &   \\
207.9 & 0.9281 &   & X  &   \\
217.9 & 0.9639 &   & X  &   \\
220.0& 0.9821 &    & X  &   \\
\end{tabular}
\end{center}
\caption{The vertical locations of the thermistors in the 10 columns defined in Table~\ref{tab:location}. The ``X" for a given column indicates that that column contained a thermistor at that vertical position.}
\label{tab:location2}
\end{table}

For the present experiment we used the 54 thermistors listed in Table~\ref{tab:location}. They were positioned in 8 columns at various radial and azimuthal locations within the sample. As can be seen, version HPCF-IIf [which was used by \cite{ABFGHLSV12}] contained only columns $V_0$ and $V_1$, both using thermistors T$_1$. In version HPCF-IIg (which was used for the present work) eight new columns, using thermistors T$_2$, were added at several different radial locations. For each column the vertical positions at which a thermistor was present are shown in Table~\ref{tab:location2}. There one sees that all columns were located in the lower half of the samples, except for $V_{0}^\prime$ and $V_{1}^\prime$ which were positioned in the top half so that the expected symmetry about the horizontal mid plane could be investigated. 

The experimental results for the logarithmic profiles reported previously by \cite{ABFGHLSV12} were obtained with HPCF-IIf which contained only the two columns $V_{0}$ and $V_{1}$, located in the lower half of the sample and each  consisting of eight T$_1$ thermistors.

\subsubsection{Thermistor calibration and method of use}
\label{sec:calib}

All 52 thermistors in the sample interior were calibrated inside the apparatus against the top- and bottom-plate thermistors \cite[]{AFB09}, which in turn had been calibrated separately against a platinum thermometer. This was done by establishing a steady state typically with an applied temperature difference of 0.8K and a mean temperature $T_m$ close to 21.50$^\circ$C. The small temperature difference was needed since the system would not reach a steady state in the absence of it, even in a time interval of a couple of days. Using parameters of the vertical logarithmic temperature profiles that were measured at various pressures and with large $\Delta T$, a small correction to the calibrations  which depended on the vertical and radial location of each of the 54 thermometers was made. 

Since all temperatures to be measured were close to $T_m$, we used the measured resistances $\Omega_0(T_m)$ and a standard value of $\Omega^{-1}(d\Omega/dT) = 0.0456 K^{-1}$ to calculate all temperatures.

\section{Results for the Classical State}
\label{sec:classical}

\begin{figure}
\centerline{\includegraphics[width=0.95 \textwidth]{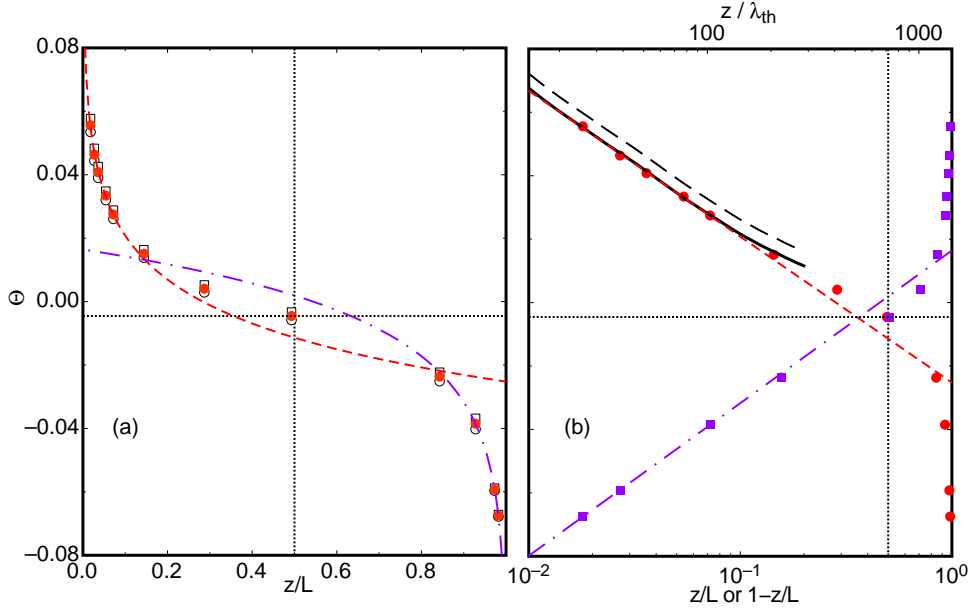}}
\caption{The dimensionless temperature $\Theta(z/L) = [T(z/L) - T_m]/\Delta T$ as a function of $z/L$ or $1-z/L$ for $\xi = 0.0180$ from run 1209041 for $\Ra = 1.99\times 10^{12}$ and $\Pra = 0.786$ (results from run 1209051 at the same \Ra\  and \Pra\  were, within the resolution of the figure, virtually indistinguishable from these data). Open circles: $\Theta_0$. Open squares: $\Theta_1$. Solid circles: $\Theta = (\Theta_0 + \Theta_1)/2$. (a) gives the data on linear scales (as a function of $z$ only) while (b) gives them on semilogarithmic scales. The dashed and dash-dotted lines are fits of Eq.~\ref{eq:Theta2} and \ref{eq:Theta3} to the data for $z/L < 0.08$ and $1-z/L < 0.08$ respectively. They correspond to $A = -0.0200\pm 0.0005$, $B = -0.0206\pm 0.0015$, $A^\prime = 0.0209$, and $B^\prime = 0.0210$. The vertical dotted line is located at $z/L = 1/2$ and the horizontal dotted line is at $\Theta(z/L=0.5) = \phi = -0.0046$.}
\label{fig:log_profile2}
\end{figure}

\subsection{Results for $\Ra = 2\times 10^{12}$}
\label{sec:2e12}

We chose $\Ra = 2\times 10^{12}$ for a detailed examination of the data because this is the largest \Ra\ at which comparison with DNS by \cite{SLV11} for $\Pra = 0.7$ [see \cite{ABFGHLSV12} for the results of an analysis of these data] is possible; but results at other \Ra\ in the classical range $6\times 10^{11} \alt \Ra \alt 10^{13}$ covered by our measurements are similar.

\subsubsection{Axial dependence: The logarithmic temperature profiles}

In Fig.~\ref{fig:log_profile2}(a) and (b) we show results for $\Theta$ from run 1209041 ($P = 2.065$ bars, $T_m = 21.5$ K, $\Ra = 1.99\times 10^{12}$, $\Pra = 0.786$) as a function of $z/L$ or $1 - z/L$ at the radial position $\xi = (R-r)/R = 0.018$. 
The open circles in (a), to be referred to as $\Theta_0$,  are from thermistor columns $V_{0}$ and $V_{0}^\prime$, while the open squares ($\Theta_1$) are from columns $V_{1}$ and $V_{1}^\prime$. As seen from Table~\ref{tab:location}, these two column pairs are located at the same radial position but at azimuthal positions $\phi$ that differ by $\pi$. Over the measuring positions from
$z/L = 0.018$ to 0.982, 
the vertical time-averaged temperature $T(z/L)$ 
only changed by 13 percent of $\Delta T$. This indicates that
the measured temperature profile is well inside the bulk, outside the thermal
boundary layers.  There is a small systematic difference between the temperatures at the two azimuthal locations. At the horizontal mid plane, where we define $\phi \equiv \Theta(z = L/2)$, this difference corresponds to $\phi_{0} - \phi_{1} = - 0.0025$. With $\Delta T = 9.00$K for this run, this indicates a temperature difference of -0.022K. We attribute this difference to a dependence on azimuthal position of the time averaged temperatures along vertical lines near the side wall due to the preferred orientation of a weak remnant of a   large-scale circulation (LSC). For all subsequent analyses we shall use the average $\Theta \equiv (\Theta_0 + \Theta_1)/2$ of the two columns which is shown as solid circles (red online) as a function of $z/L$ in Figs.~\ref{fig:log_profile2} (a) and  (b).
 
\subsubsection{Non-Bousinesq effects and parameter estimates}
\label{sec:NOB}

Before proceeding further we first show that non-Oberbeck-Boussinesq (non-OB) effects do not have an undue influence on the remainder of our analysis. 

For the OB case we expect that $\phi = (\phi_0 + \phi_1)/2 = 0$. One sees that there is a slight offset of $\Theta$ at $z/L = 0.5$, corresponding to $\phi = - 0.0046$ or $T_c - T_m = -0.041$K ($T_c$ is the temperature in the horizontal mid plane of the sample which, as we shall see below, is nearly independent of the radial position $r$). We regard this as indicative of small deviations of the physical system from the OB approximation \cite[]{Ob79,Bo03}, as discussed in a number of publications (see, for instance, \cite{WL91,ZCL97,ABFFGL06,AFFGL07,ACFFGLS08}). The result for $\phi$ has the same sign as and a magnitude consistent with those for compressed ethane gas (not too close to the critical point) reported by \cite{AFFGL07}. Although we are not aware of a rigorous way to correct for this shift of $T_c$, we shall consider (when necessary) temperature profiles shifted by $\phi$ and believe that this correction is adequate in view of the small value of $\phi$.

For the Boussinesq system we expect that the amplitudes $A$ and $A^\prime$ (see Eq.~\ref{eq:Theta2} and \ref{eq:Theta3}) should have the same absolute values, {\it i.e.} $-A = A^\prime$. For the case shown in Fig.~\ref{fig:log_profile2}(b) we have $A = -0.0200\pm 0.0005$ and $A^\prime = 0.0209$ (it is not possible to determine independently a reliable uncertainty for $A^\prime$ and $B^\prime$ because the fit is based on only three data points, but we expect these uncertainties to be about the same as those for $A$ and $B$).  One sees that, within expected uncertainties, $A = -A^\prime$, consistent with the OB case. Below, in Sec.~\ref{sec:para_on_Ra}, we shall examine $A/A^\prime$ as a function of \Ra\ and see that, averaged over all \Ra\ of this study, $A/A^\prime = -0.989 \pm 0.008$ where the uncertainty is the 67\% confidence limit. Thus we find no conclusive evidence for non-OB effects on $A/A^\prime$. As one might have expected, this implies that the non-OB conditions lead to unequal temperature drops across the two thermal BLs adjacent to the top and bottom plate which in turn result in a shift of $T_c$ relative to $T_m$; however, aside from this temperature shift (which is uniform throughout the bulk) the reflection symmetry about the horizontal mid plane is maintained throughout the bulk of the sample.  

In order to assess more generally the possible importance of non-OB effects, we consider the dimensionless parameter  $\alpha \Delta T$, which is frequently used as an empirical measure of the size of these effects. For our sample it had the value 0.033. In the work of \cite{UHKMSS12} for instance $\alpha \Delta T$ reached values as large as 0.38. We note also that our value of $\phi$ is an order of magnitude smaller than the largest encountered by \cite{UHKMSS12}; thus, as discussed in more detail by \cite{HFNBA13},  there is no need for concerns such as those expressed by  \cite{UHKMSS12} about an excessive influence of non-OB effects on our results. 

\subsubsection{Comparison with direct numerical simulation}
\label{sec:DNS}

The DNS results of \cite{SLV11} [see Fig. 3 of \cite{ABFGHLSV12}] for $\Ra = 2\times 10^{12}$, $\Pra = 0.7$, and $\xi =0.0180$ are shown as a dashed line in Fig.~\ref{fig:log_profile2}(b). These results are an azimuthal average of the computed temperature field for the OB system. For them $\phi = 0$ and thus they are offset vertically relative to the experimental data. In order to allow a better comparison with the experiment, we shifted them by a constant increment of $\phi = -0.0046$ as found from the experiment. This yielded the solid line in Fig.~\ref{fig:log_profile2}(b), which over its range of reliability agrees superbly with the measurements. A fit of Eq.~\ref{eq:Theta2} to the DNS data over the range $0.018 \leq z/L \leq 0.072$  used in the fit to the experimental data gave $A = -0.0201$, in quantitative agreement with the measurements.

\subsubsection{Deviations from the logarithmic dependence}
\label{sec:devlog}

Here we consider the deviations of $\Theta(z/L)$ from the fits $\Theta_{fit}$ of Eqs.~\ref{eq:Theta2} or \ref{eq:Theta3} to the five data points with $z/L < 0.08$ or the three points with $1-z/L < 0.08$ when the distances from the plates become larger and approach the horizontal mid plane at $z/L = 1-z/L = 0.5$. 

From Fig.~\ref{fig:log_profile2} and a close examination of the numerical data one sees that these extrapolations reach the mid-plane temperature $\phi$ at $z/L = Z_0 = 0.36$ and $Z_0^\prime = 1-z/L = 0.36$. The symmetry revealed by the finding that $Z_0 = Z_0^\prime$, combined with the symmetry $-A=A^\prime$ already noted in Sec.~\ref{sec:NOB}, shows that the non-OB effects of significant size are limited to the boundary layers (where they cause a uniform shift of the bulk temperature field) and do not have any other measurable influence on the bulk. Based on different measurements for a sample with $\Pra = 4.4$ and 5.5 a similar conclusion was reached before \cite[]{BA07_EPL}. This is of course as expected, since temperature variations in the bulk are relatively small and thus fluid properties are nearly constant.

From Eqs.~\ref{eq:Theta2} and \ref{eq:Theta3}, and our assumption that non-OB effects can be adequately compensated by a shift of the temperatures $\Theta(z/L)$ by $\phi$, one has the relations
\be
(B - \phi)/A = -ln(Z_0)
\label{eq:B_over_A}
\ee 
and
\be
(B^\prime - \phi)/A^\prime = -ln(Z_0^\prime)\ .
\label{eq:Bprime_over_Aprime}
\ee 
If the logarithmic profiles extended all the way to the horizontal mid plane, then we would have $Z_0 = Z_0^\prime = 0.5$ and $(B - \phi)/A = (B^\prime - \phi)/A^\prime = -ln(0.5) \simeq 0.69$. The experimental finding that $Z_0 = Z_0^\prime < 0.5$ is an indicator of the existence of an outer layer where deviations from the logarithmic profiles occur. 

The parameters $(B - \phi)/A$ and $(B^\prime - \phi)/A^\prime$ are convenient indicators (but not a direct measure) of the widths of the logarithmic and the outer layers which can be derived directly from the parameters of the logarithmic fits to the data. Thus, below in Fig.~\ref{fig:log_profile10}(a), we show results for  them as a function of \Ra. The data indicate that these  widths are independent of \Ra\ within the resolution of the measurements. They yield average values $(B - \phi)/A = 0.967\pm 0.023$ and $(B^\prime - \phi)/A^\prime = 0.953 \pm 0.022$ (corresponding to $Z_0 = 0.380$ and $Z_0^\prime = 0.386$), indicating that the outer layer remains symmetric about the mid plane and of the same width independent of \Ra. 

\begin{figure}
\begin{center}
\includegraphics[width=0.8\textwidth]{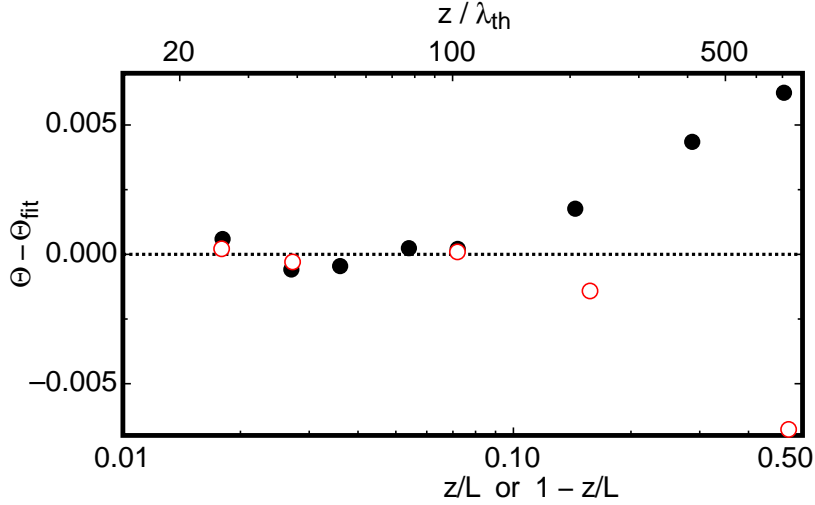}
\caption{Deviations of the data in Fig.~\ref{fig:log_profile2} ($\Ra = 1.99\times10^{12}$) from a fit of Eq.~\ref{eq:Theta2} to the points for $z/L < 0.1$ (solid circles) or Eq.~\ref{eq:Theta3} to the points with $1 - z/L < 0.1$ (open circles) as a function of $z/L$ or $1-z/L$ and of $z/\lambda_{th}$ on a logarithmic scale. For this \Ra\ the measurements yielded $\Nu = 716$, giving $\lambda_{th}/L = 1/(2\Nu) = 7.0\times 10^{-4}$.}
\label{fig:deviations2}
\end{center}
\end{figure}

Next we examine more directly the deviations of $\Theta(z/L)$ from $\Theta_{fit}$. As $z/L$ or $1-z/L$ exceed 0.1, one can see from Fig.~\ref{fig:log_profile2} that deviations occur. This is more apparent in the plot of $\Theta - \Theta_{fit}$ as a function of $z/L$ shown in Fig.~\ref{fig:deviations2}. For $z/L \alt 0.1$ the rms deviations are less than about  $5\times 10^{-4}$, {\it i.e.} 0.05\% of $\Delta T \simeq 9$K or about 4 mK. For larger $z/L$ they increase, reaching values near 0.006 (or 0.6\% of $\Delta T$ or 54 mK) at the sample mid plane. The existence of a log layer and an outer layer now can be seen clearly. This is analogous to results for various shear flows [see {\it e.g.} \cite{Po00}]; this analogy will be discussed in detail in Sec.~\ref{sec:compLoW}. 

\subsubsection{Radial dependence of $\Theta$} 
\label{sec:rad}

\begin{figure}
\begin{center}
\includegraphics[width=0.95\textwidth]{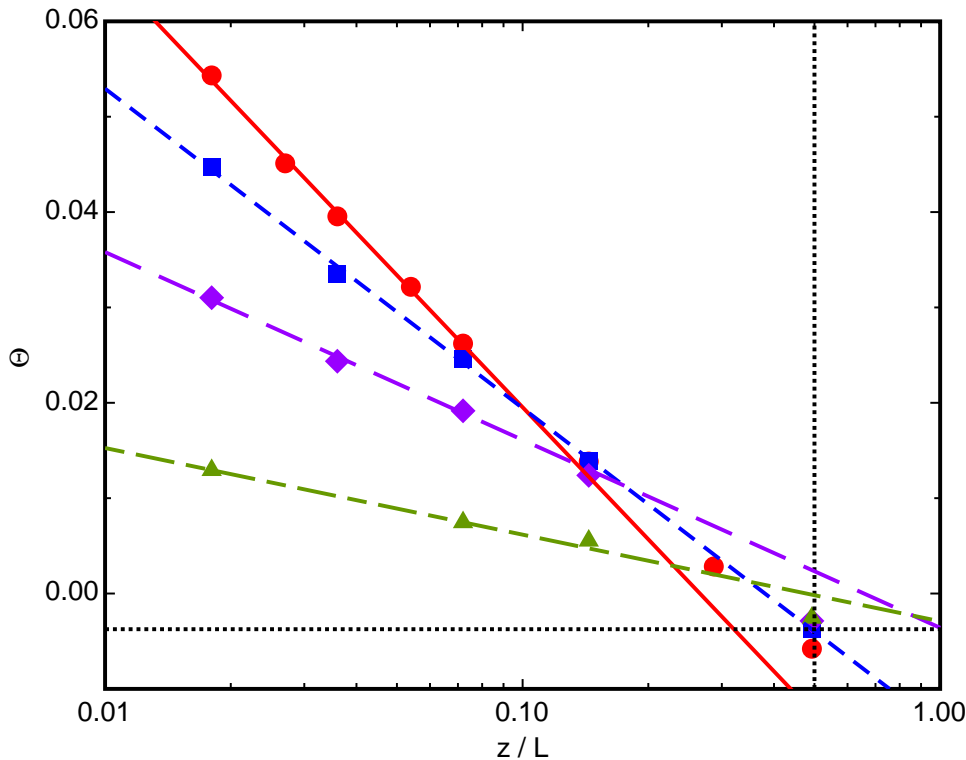}
\caption{Results at the four radial location $(R-r)/L$ for $\Theta(z)$ from run 1209041 ($\Ra = 1.99\times 10^{12}, \Pra = 0.786$) from HPCF-IIg. Near the left edge of the graph we have, from top to bottom, $\xi = (R-r)/R = 0.0180, 0.0356, 0.0716$, and 0.2680. Note that all four data sets have very nearly the same value at $z/L = 0.5$, showing that the horizontal mid-plane temperature $\phi \equiv \Theta(z/L=0.5)$ is nearly independent of $r$. The averaged measured value $\langle \phi\rangle = -0.0037$ is shown as a horizontal dotted line. The vertical dotted line is at $z/L = 0.5$, corresponding to the horizontal mid plane.}
\label{fig:radial1}
\end{center}
\end{figure}

\begin{figure}
\begin{center}
\includegraphics[width=0.95\textwidth]{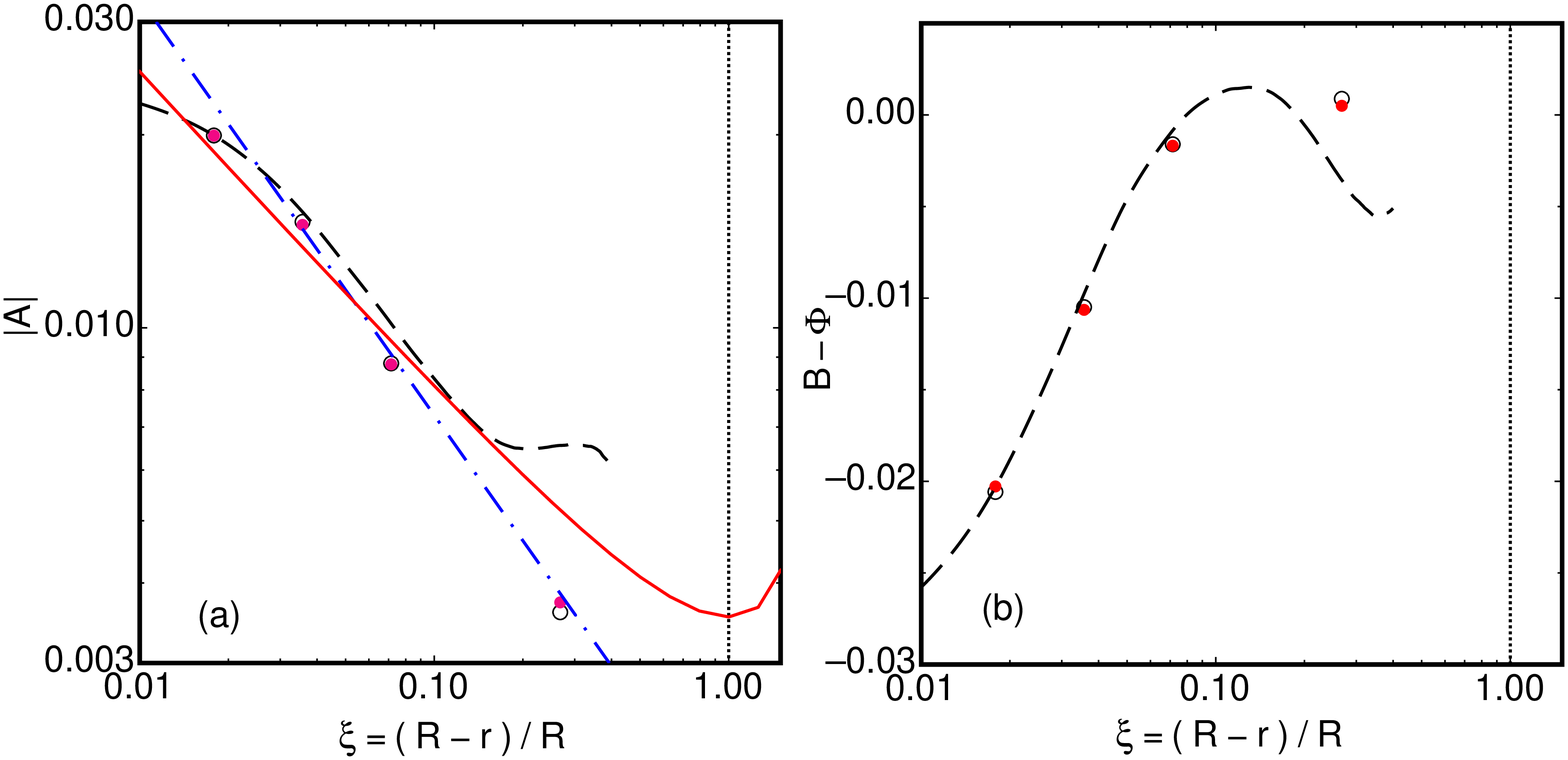}
\caption{The parameters $|A|$ and $B - \phi$ of the logarithmic temperature profiles for 
$\Ra = 2.0\times 10^{12}$ and $\Pra \simeq 0.787$.  Open circles: run 1209041. Solid circle: run 1209051. Dashed line: DNS by \cite{SLV11} [see \cite{ABFGHLSV12}] for $\Ra = 2\times 10^{12}$ and $\Pra = 0.7$.
Dash-dotted line (blue online): $|A| = 0.0016*\xi^{-0.65}$ with $\xi = (R-r)/R$. 
Solid line (red online): Eq. \ref{eq:GL}  with $A_1 = 0.00354$. The vertical dotted line is located at the axis of the sample.}
\label{fig:radial2}
\end{center}
\end{figure}

An important aspect of RBC is the existence of an additional coordinate that does not exist in shear flow. In RBC the parameters of the log profiles can, and indeed do, vary in the radial direction. This was observed first in DNS [\cite{SLV11}; see \cite{ABFGHLSV12}], and is illustrated in Fig.~\ref{fig:radial1} which shows experimental data for $\Theta$ as a function of $z/L$ for four radial positions and $\Ra = 1.99\times 10^{12}$. One sees that the slopes of the lines fitted to the points (on a semi-log scale) decrease as the sample centerline [$\xi = (R-r)/R = 1$] is approached. At the horizontal mid plane of the sample  ($z/L = 0.5$) the data at all radial positions $\xi$ nearly coincide, showing that there is very little variation of the temperature in that plane. 

Figure~\ref{fig:radial2} shows the parameters $A$ and $B-\phi$ for two separate runs at $\Ra = 1.99\times 10^{12}$ as a function of $\xi$. The two data sets are in near-perfect agreement. Here the mid-plane temperature-shift $\phi = \Theta(z/L=0.5) = (T_m - T_c)/\Delta T$ was subtracted from $B$ so as to compensate for the non-OB effects. One sees that the magnitudes of $A$ and of $B-\phi$ decrease as the radial position decreases. This result is not surprising if one assumes that the logarithmic profile of $\Theta$ is the result of the plumes that are emitted from the thermal boundary layers adjacent to the plates and travel upward or downward through the sample. Plumes tend to rise or fall where there is an up flow or down flow of a large-scale circulation. Thus, for a cylindrical sample of aspect ratio near unity plumes are found in greater abundance near the side walls.

Also shown in Fig.~\ref{fig:radial2}, as a dashed line, is the DNS result for $\Ra = 2\times 10^{12}, \Pra = 0.7$ [\cite{SLV11}; see \cite{ABFGHLSV12}]. From the DNS, azimuthal temperature averages were computed in order to increase the precision of the results. For small $(R-r)/R$ the agreement with the data is excellent; it deteriorates slightly as the sample center-line at $r = 0$ is approached, probably because the accuracy of the DNS decreases because the azimuthal averaging includes fewer grid points at the smaller radii.

Recently \cite{GL12} examined the logarithmic profiles in the ultimate state and for that case proposed that the amplitude should vary as
\be
|A| = \frac{A_1}{\sqrt{2\xi - \xi^2}}\ .
\label{eq:GL}
\ee
Near the side wall (but inside the boundary layers) one then has $|A(\xi)| \sim (A_1/\sqrt{2})\xi^{-1/2}$, and at the vertical center line $|A(\xi = 1)| = A_1$. Although the prediction was not for the classical state illustrated by Fig.~\ref{fig:radial2}, it may be relevant regardless of the mechanism involved in the generation of the logarithmic profiles because it is based on geometric arguments that assume a circular path of the large-scale circulation. Thus we compare it with the data by showing it as the solid line (red online) in the figure. Here the coefficient $A_1$ was adjusted to match the points at small $\xi$ which yielded $A_1 = 0.0035$. The data fall more rapidly with increasing $\xi$ and the $A_1$ estimate must be high. An alternative representation of  the data is provided by the dot-dashed line (blue online) which is the power law $|A| = A_0*\xi^{-0.65}$. \cite{GL12} had indicated (for the ultimate state) that for samples like ours with an aspect ratio less than one an exponent more negative than the value -1/2 indicated by Eq.~\ref{eq:GL} may be appropriate because the streamlines of the large-scale circulation will deviate more from a circle. The power law extrapolates to $A_0 = 0.0016$ at $\xi = 1$, but this extrapolation can not be valid either and must yield a low value because it leads to a singularity in $A(r)$ at $\xi = 1$. An extrapolation that is rounded and has zero slope at the vertical center line would yield a larger value, somewhere between 0.002 and 0.003.

\begin{figure}
\centerline{\includegraphics[width=0.85 \textwidth]{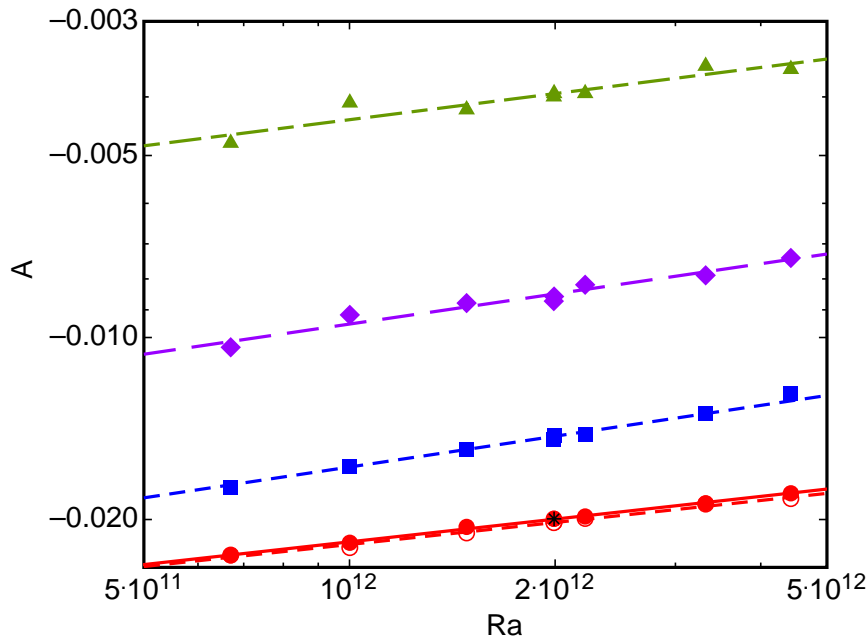}}
\caption{The amplitudes $A$ and $A^\prime$ of the logarithmic dependence of $\Theta$ on $z/L$ for the classical state as a function of the Rayleigh number \Ra.  From bottom to top, the data are for the radial positions $\xi = 0.0180$,  0.0356, 0.0716, and 0.268 (see Tables~\ref{tab:location} and \ref{tab:location2}). The lines are fits of Eq.~\ref{eq:A} or \ref{eq:Aprime}  to the data. All data are for $A$, except that for $\xi =0.018$ the open circles (several of them hidden by the solid circles) are for $-A^\prime$. The black star corresponds to the data in Fig.~\ref{fig:log_profile2}.}
\label{fig:AofRa}
\end{figure}

\subsection{Dependence of the parameters of the logarithmic term on \Ra}
\label{sec:para_on_Ra}

In Fig.~\ref{fig:AofRa} we show the amplitude $A$ as a function of \Ra\ on double logarithmic scales for the four values of $\xi$. One sees that $A(\xi,\Ra)$ depends on \Ra\ at all radial positions, and fits of the power law 
\be
A = A_0(\xi) \Ra^{-\eta}
\label{eq:A}
\ee
 to the data for $z/L < 0.08$ or
 \be
 A^\prime = A_0^\prime(\xi) \Ra^{-\eta^\prime}
 \label{eq:Aprime}
 \ee
 to the data for $1-z/L < 0.08$ (the lines in the figure) give similar exponent values. 

The dependences of $A$ and $A^\prime$ on \Ra\ are not very strong, and thus accurate data are needed to determine $\eta$ and $\eta^\prime$   with meaningful accuracy. As can be seen from Fig.~\ref{fig:radial1}, only the data for $\xi =0.018$ are based on fits to five temperatures for $z/L < 0.08$; for all other $\xi$ there were only two or three measurements and thus systematic errors arising from errors in the thermometer locations and calibrations are expected to be larger. In addition, we expect errors of the thermometer locations to increase as $\xi$ increases because the thermistors are suspended on ever increasing lengths of ceramic rods (see Sec.~\ref{sec:types}) and thus more difficult to position with high accuracy. Further, the size of the logarithmic amplitudes decreases strongly with increasing $\xi$, and thus the random errors of the measurements become more important. For these reasons we  believe that only the measurements for $\xi = 0.018$ (open and solid circles in Fig.~\ref{fig:AofRa}) warrant a quantitative study of the \Ra\  dependence of the  coefficients. 
A fit of Eqs.~\ref{eq:A} and \ref{eq:Aprime} to the data for $\xi = 0.018$ yielded 
\be
A_0 = -0.686 \pm 0.06;\ \ \ \eta = 0.125 \pm 0.003
\label{eq:A0}
\ee
and
\be
A_0^\prime = 0.623 \pm 0.05;\ \ \ \eta^\prime = 0.121 \pm 0.003\ .
\label{eq:A0prime}
\ee
These results are consistent with $\eta = \eta^\prime = 0.123$ with an uncertainty of 0.01 or smaller. A fit to the data with $\eta$ and $\eta^\prime$ fixed at 0.123 gave $A = -0.651 \pm 0.001$ and $A^\prime = 0.659 \pm 0.002$, corresponding to the ratio $A/A^\prime = -0.988 \pm 0.005$. This result is consistent with the value $A/A^\prime = -0.989 \pm 0.008$  given in Sec.~\ref{sec:NOB}   and based on an analysis of the data at a single value of \Ra. It suggests a value slightly less than one for $-A/A^\prime$; but since all error estimates are 67\% confidence limits we can not really assert that the ratio differs significantly from unity.

\begin{figure}
\centerline{\includegraphics[width=0.85 \textwidth]{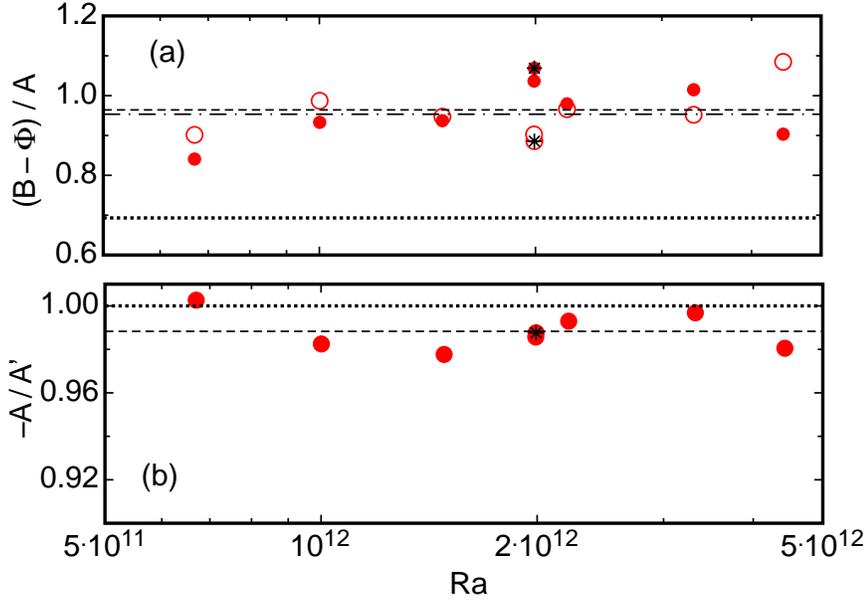}}
\caption{The amplitude ratio $A/A^\prime$ and the parameter ratio $(B-\phi)/A$ of the logarithmic dependence of $\Theta$ on $z/L$ for the classical state as a function of the Rayleigh number \Ra. In (a) the open  and solid circles are for $(B^\prime - \phi)/A^\prime$  and $(B-\phi)/A$ respectively. The dotted horizontal lines correspond to the Boussinesq values $-A/A^\prime = 1$ and $(B-\phi)/A = - \ln(0.5) = 0.693$. The short-dashed and dash-dotted lines in (a) correspond to the averaged values $\langle (B-\phi)/A \rangle = 0.964$ and $\langle (B^\prime-\phi)/A^\prime \rangle = 0.953$ respectively.  The
short-dashed line in (b) corresponds to the averaged value $\langle -A/A^\prime \rangle = 0.989$. The black stars correspond to the data in Fig.~\ref{fig:log_profile2}.}
\label{fig:log_profile10}
\end{figure}

The dependence of $A$ on \Ra\ differs from shear flow, where the logarithmic profile of the time averaged stream wise velocity (when scaled appropriately by the friction velocity)  has an amplitude $1/\kappa$ where $\kappa$ is the von K\'arm\'an constant and independent of the Reynolds number. As discussed below in Sec.~\ref{sec:compLoW}, the radial dependence of $A$ does not have an analog in the shear flow case as the radial dimension is unique to RBC. 

We note that, for the ultimate state, \cite{GL12} had predicted that $A$ should follow a power law as \Ra\ is varied with an effective exponent $\eta_{GL} \simeq 0.043$, but to our knowledge there is no reason for this result to be relevant to the classical state. In any case, it differs from the experimental value. A recent analysis based on a multi-layer model \cite[]{SCCZBH14} arrived at a log layer with an amplitude of the logarithm varying with \Ra\ as $\Ra^{-0.162}$ (this model does to seem to distinguish between the classical and the ultimate state) which is slightly higher than the experimental value.

For log profiles that extend all the way to the horizontal mid plane  ($Z_0 = Z_0^\prime = 1/2$) we expect (see Eq.~\ref{eq:B_over_A}) $(B - \phi)/A = (B^\prime - \phi)/A^\prime = -ln(0.5) \simeq 0.69$ independent of \Ra. We show both $-A/A^\prime$ and $(B - \phi)/A$ in Fig.~\ref{fig:log_profile10}. As already noted above,  $-A/A^\prime$ is very close to one and within the resolution independent of \Ra. However, $(B - \phi)/A$ is larger than 0.69. The data yield $\langle (B - \phi)/A \rangle = 0.967 \pm 0.023$ ($Z_0 = 0.380 \pm 0.009$) and  $\langle (B^\prime - \phi)/A^\prime \rangle = 0.953 \pm 0.022$ ($Z_0^\prime = 0.385\pm 0.008$). As already discussed in Sec.~\ref{sec:devlog}, we attribute this to the existence of an outer layer between each log layer and the mid plane. Within their resolution the data do not reveal any \Ra\ dependence of the width of the log layer, and the result $Z_0 = Z_0^\prime$ indicates that the outer layer is symmetrically distributed about the mid plane.

\subsection{Comparison with the ``Law of the wall" for shear flows}
\label{sec:compLoW}

\begin{figure}
\centerline{\includegraphics[width=0.85 \textwidth]{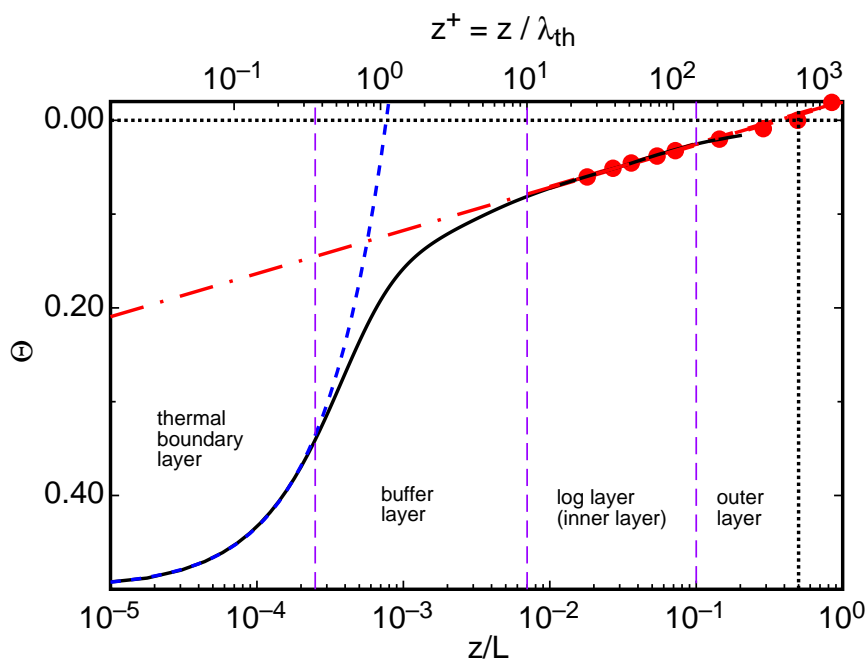}}
\caption{The dimensionless temperature $\Theta$ as a function of $z/L$ or $z/\lambda_{th}$ for $\Ra = 2\times10^{12}$ and $\xi = 0.018$. Solid circles (red online): run 1209041 (the same data as those shown in Fig.~\ref{fig:log_profile2} and \ref{fig:deviations2}). Dash-dotted line (red online): Fit of Eq.~\ref{eq:Theta2} to the data for $z/L < 0.08$. Solid black line: DNS for $\Ra = 2\times 10^{12}$ from Fig. 3 of \cite{ABFGHLSV12}. Short dashed line (blue online): Fit of $\Theta = \Theta_0 * (z/L)$ to the DNS data for $z/L \leq 10^{-4}$ which gave $\Theta_0 = -670$. The vertical longer-dashed lines (purple online) show the approximate locations of the boundaries between four different regions, in analogy to those for the velocity in shear flow.}
\label{fig:analogy}
\end{figure}

Using Fig.~\ref{fig:analogy},  we call attention to the close similarities and interesting differences between the time-averaged temperature-profiles in RBC and the time-averaged velocity-profiles in high-Reynolds-number  shear flows. To this end the scale of $\Theta$ is increasing in the downward direction, thus emphasizing the similarity to the shape of the down-stream shear-flow velocity as a function of distance from the wall [see {e.g.} \cite{Po00}]. The solid line in the figure is the result of DNS for $\Ra = 2\times 10^{12}$ and $\Pra = 0.7$ [\cite{SLV11} as reported by \cite{ABFGHLSV12}]. The solid circles are results of the present work. 

We show $\Theta$ {\it vs.} $z/L$ (bottom scale) and $z^+ \equiv z/\lambda_{th}$ (top scale). Here $\lambda_{th}$ is taken to be the thermal boundary-layer thickness $\lambda_{th} \equiv L/(2 \Nu)$, which corresponds to the vertical position of the extrapolation of the conduction profile near the plate to the horizontal mid plane of the sample. With $\Nu = 716$ from the present experiment, we find $\lambda_{th}/L \simeq 7.0\times 10^{-4}$ or $\lambda_{th} = 1.56$ mm for the particular case shown here. We suggest that it is reasonable that $\lambda_{th}$ should be the appropriate inner scale for calculating the distance from the plate in ``wall units" because the excitations in RBC presumably are plumes emanating from the thermal boundary layers. Plumes, when they first form,  are believed to be one-dimensional excitations (horizontal lines), not  unlike isolated convection rolls, that form in the marginally stable laminar (albeit time dependent) thermal boundary layer (see Sec.~3.2.1 of \cite{FBA08}, \cite{BWA12}, and references therein). Thus the characteristic width of the plumes is expected to be of the order of $\lambda_{th}$. Assuming that the plumes play a role analogous to the velocity excitations, or ``coherent eddies" [see, {\it e.g.} \cite{To76,PC82,PHC86}], in shear flow, the use of $\lambda_{th}$ would be analogous to using the thickness of the viscous sublayer $\delta_\nu$ in shear flows (which determines the ``size" of the coherent eddies when they are first formed) to obtain the distance $y^+ \equiv y/\delta_\nu$ from the wall in ``wall units" [see. {\it e.g.} \cite{Po00}]. 

Looking at Fig.~\ref{fig:analogy}, one sees that there is a thermal boundary layer for $z^+ \alt 0.4$ where the temperature profile is well approximated by a linear function $\Theta \propto z^+$ (the short dashed line). We take it to play a role analogous to that of the ``viscous sublayer" in shear flow where the velocity varies linearly with $y^+$; but this sublayer is thicker, extending to $y^+ \simeq 3$ or 4. This difference, we assume, is due to the fact that the viscous sublayer is turbulent, while the thermal boundary layer in the classical state of RBC, although time dependent,  has a profile close to the Prandtl-Blasius laminar profile when  the time dependence is properly taken into account  \cite[]{ZX10,SZGVXL12}. Thus, in view of the definition of $\lambda_{th}$, the thermal boundary layer has a thickness less than one.

Beyond the thermal BL there is a ``buffer layer" which provides the crossover from the linear dependence of $\Theta$ on $z^+$ to the logarithmic dependence. The buffer layer extends to $z^+ \simeq 10$, which is only somewhat smaller than  the value of $y^+ \simeq 25$ or so at the upper limit of the shear-flow buffer layer.

Over the range $10 \alt z^+ \alt 150$ the log layer is found in the RBC system. It is difficult to say precisely where the log layer ends in RBC because the deviations from it are very small all the way up to the sample center (see Sec.~\ref{sec:devlog} and Fig.~\ref{fig:deviations2}). Thus, we can only say that the log-layer range also corresponds approximately to the range of $y^+$ over which the log layer occurs in shear flows [see, {\it e.g.} \cite{WW89,Po00}].

 In the above picture, although there are similarities between RBC and shear flow, there is a significant difference. In the classical state of RBC the excitations form in a laminar (although time dependent) layer; in shear flows at large Reynolds numbers there is no laminar viscous boundary layer and instead the eddies form in a turbulent viscous sublayer which exists only because of the boundary condition of vanishing velocity at a rigid wall. Regardless of this difference, a picture evolves in which the consequences are analogous: The excitations (eddies or plumes) become ejected through a buffer layer and become responsible for the formation of a  log layer. 

For $z^+ \agt 150$ there is an ``outer layer" where there are small deviations from the log profile which are barely noticeable in Fig.~\ref{fig:analogy} but clearly see in Fig.~\ref{fig:deviations2}, analogous to the outer layer in shear flow. It is interesting to note that recent measurements of the spectra of fluctuations in this system \cite[]{HGBA13} showed that the same analogy extends also to the dynamics. The data revealed that the fluctuation power varies with the frequency $f$ as $f^{-1}$ over about one decade  of $f$ when $z$ is in the log-layer range of the variance of the fluctuating temperature. In contrast,  an $f$-dependence with a more negative exponent, closer to $f^{-1.5}$, was found  in an outer layer; this is not unlike the Kolmogorov scaling $f^{-1.7}$ found for frequency spectra in an outer layer near the center of pipe flow \cite[]{RHVBS13}. Thus, we expect that the range $z^+ \agt 150$ does indeed correspond to an outer layer similar to that found in shear flows.

Finally we note an important difference between shear flows and RBC. In fully developed turbulent shear flows the system is homogeneous in the downstream (``stream wise") direction, with significant variations of the turbulent flow properties only in the transverse ($y$, ``spanwise") direction. RBC is more complex because its properties vary in two dimensions, namely in the vertical ($z$) direction (corresponding to the transverse shear-flow direction) and in the horizontal ($r$) direction which has no analog in shear flow. Thus, as we saw in Sec.~\ref{sec:rad}, the parameters $A(r)$ and $B(r)$ of the logarithmic profiles are not constant in the horizontal plane but rather (in our cylindrical geometry) depend on the radial position. This is consistent with the experimental observation, as well as with findings from DNS,  that plumes tend to rise (fall) mostly in upflow (downflow) regions of the large-scale circulation, with an associated plume density that is dependent on the radial position within the sample. 
%It would be interesting to examine the radial variation of the log profiles in samples where the LSC is suppressed, for instance by baffles or screens \cite[]{CCS96,XL97,XQ99}. In such a system the plumes rise vertically at all radial positions and not only near the side wall. We would then expect $A$ and $A^\prime$ to be less dependent on $r$.

Another important difference between RBC and shear flow is that the amplitude $A$ of the logarithmic profile (even at a fixed radial position) is found to be dependent on the Rayleigh number (see Fig.~\ref{fig:AofRa} above), while the corresponding amplitude in shear flow is independent of the Reynolds number and, when the velocity is scaled appropriately, is equal to the inverse of the universal von K\'arm\'an constant $\kappa$. However, recent as yet unpublished measurements by P. Wei and G. Ahlers of the logarithmic temperature profile in a fluid with $\Pra = 12.3$ revealed a \Ra\ independent $A$, suggesting that the \Ra\ dependence of $A$ is characteristic only of small \Pra. Clearly, regarding the \Ra\ and \Pra\ dependences of the parameters  there is much left to be understood.

\section{Results for the Ultimate State}
\label{sec:ult}

\begin{figure}
\centerline{\includegraphics[width=0.95 \textwidth]{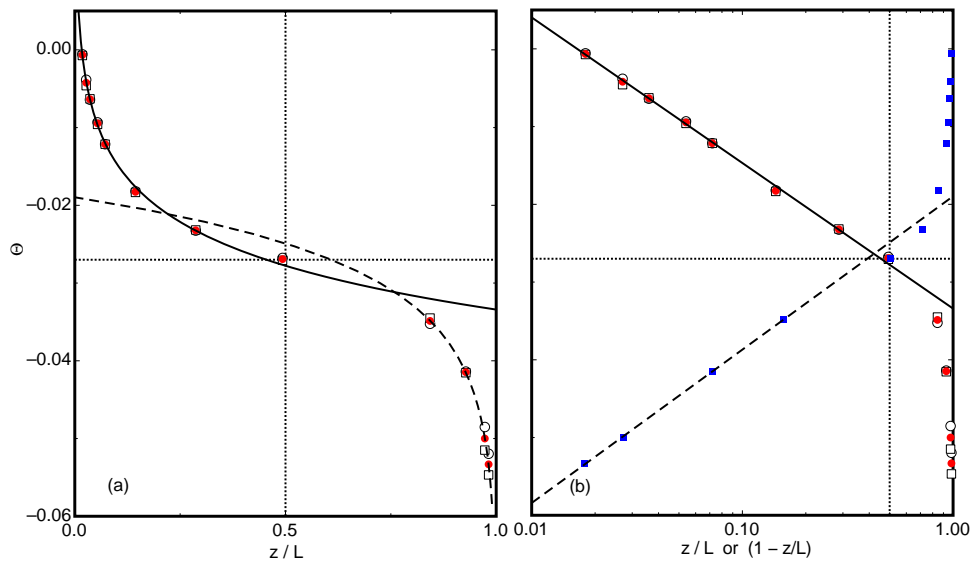}}
\caption{The dimensionless temperature $\Theta(z/L)$ as a function of $z/L$ or $(1 - z/L)$ for $\xi = 0.018$, $\Delta T = 15.22$K, $\Ra = 9.94\times 10^{14}$, and $\Pra = 0.858$ (run 1207281). The open circles (squares) are from columns $V_{0}~(V_{1})$ and $V_{0}^\prime~(V_{1}^\prime)$. The solid circles (red online) and squares (blue online) are their average. (a): the data are shown on linear scales. (b): the same data are shown on a logarithmic horizontal scale. The solid and dashed lines are fits of Eqs.~\ref{eq:Theta2} and \ref{eq:Theta3} to the data for $z/L < 0.08$ and $1-z/L < 0.08$ respectively.  
The vertical dotted line is located at $z/L = 1/2$ and the horizontal dotted line passes through the measured value of $\phi = \Theta(z/L=0.5) = -0.0270$. For this run the Nusselt number was 5540.}
\label{fig:log_profile}
\end{figure}

\subsection{Results for $\Ra = 9.9\times 10^{14}$}

\subsubsection{Axial dependence of $\Theta$: The logarithmic temperature profile} 

\begin{figure}
\begin{center}
\includegraphics[width=0.8\textwidth]{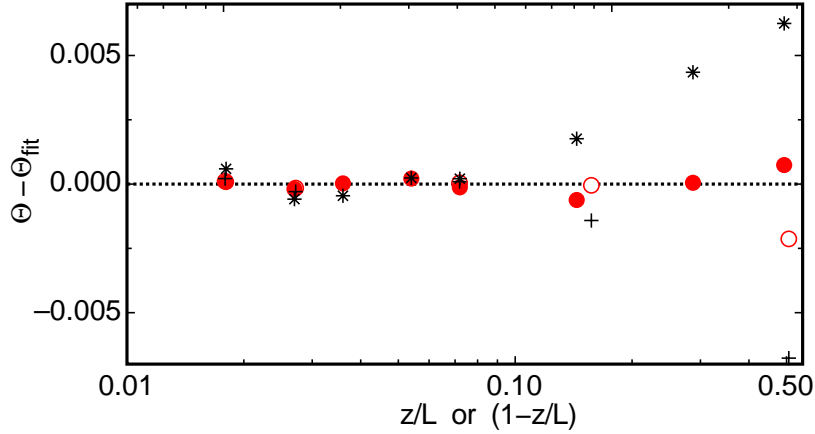}
\caption{Deviations of the solid circles in Fig.~\ref{fig:log_profile} from a fit of Eq.~\ref{eq:Theta2} to the five points for $z/L < 0.08$ (solid circles, red online) or of Eq.~\ref{eq:Theta3} to the three points for $(1-z/L) < 0.08$ (open circles, red online). Note that this figure is rendered on the same scale as Fig.~\ref{fig:deviations2} to facilitate easy comparison. Also for easy comparison we   shown the points from Fig.~\ref{fig:deviations2} as black stars and plusses.}
\label{fig:deviation}
\end{center}
\end{figure}

In Fig.~\ref{fig:log_profile} we show the reduced temperature $\Theta(z,r)$ (see Eq.~\ref{eq:Theta}) 
 at a radial position $\xi = 0.0180$ as a function of $z/L$.  As in Fig.~\ref{fig:log_profile2}, the points for $z/L \leq 0.5$ are from columns $V_{0}$ and $V_{1}$, and the points for $z/L > 0.5$ are from $V_{0}^\prime$ and $V_{1}^\prime$. For these data $\Ra = 9.94\times 10^{14}$ ({\it i.e.} the sample is in the ultimate state) and $Pr = 0.858$ ($P = 17.69$ bars, $T_m = 21.47^\circ$C, $\Delta T = 15.22$K). As seen and discussed already for the classical state, the data from $V_{0}$ and $V_{0}^\prime$ ($V_{1}$ and $V_{1}^\prime$) differ slightly from each other and we shall use their averages for further analysis.   

From Fig.~\ref{fig:log_profile}(b) it is clear that the data are described well by Eqs.~\ref{eq:Theta2} and \ref{eq:Theta3}. The solid and dashed lines in (a) and (b)  are fits of these equations to the solid circles or squares. For all fits in this paper we used only the five data points with $z/L < 0.08$ near the bottom plate and the three points with $(1-z/L) < 0.08$ near the top plate. The fits gave $A = -0.00813\pm 0.00017 $, $B = -0.0334\pm 0.0006$, $A^\prime = 0.00857$, and $B^\prime = -0.0190$ (meaningful uncertainties for $A^\prime$ and $B^\prime$ can not be established because there are not enough data points). For the amplitude ratio the data yield $-A/A^\prime = 0.95$. Although we can not reliably estimate the probable error of this ratio, we shall see below in Fig.~\ref{fig:paras}(a) that the standard deviation of many similar measurements from their mean value is 0.012.  Thus we believe that there is a significant deviation from the perfectly symmetric case $-A = A^\prime$.

In Fig.~\ref{fig:deviation} we show the deviations of $\Theta$ from the above fits as solid and open circles. Except for very close to the horizontal mid plane [$z/L = (1-z/L) = 0.5$]  the points generally fall within a fraction of a part per thousand of the fit, with a standard deviation of $2\times 10^{-4}$ corresponding to 3 mK. 
This differs from what was found for the classical state where deviations became apparent already for $z/L \agt 0.1$ and $(1-z/L) \agt 0.1$ (for comparison the classical-state data from Fig.~\ref{fig:deviations2} are shown in Fig.~\ref{fig:deviation} as  stars and as plusses). Thus, in the ultimate state we find no real evidence for an outer layer, and the log layers seems to  extend nearly to the sample center. Consistent with this, the extrapolations of the two log profiles cross the horizontal line  $\Theta = \phi$ when $Z_0 = 0.46$ and $Z_0^\prime = 0.39$, much closer to the sample center at $z/L = 0.50$ than was the case for the classical state. 

The near-absence of an ``outer layer", at least for $z/L < 0.5$, is consistent with the arguments by \cite{GL11} for the ultimate state which predict  logarithmic profiles extending throughout the sample from two very thin thermal sublayers, one each near the sample top and bottom. In the OB case the log profiles are expected to meet in the sample center without any outer layer at $z/L = 0.5$. It was suggested by S. Grossmann (private communication) that the small deviation from the log profile, primarily for $z/L > 0.5$, may be associated with non-OB effects (which caused a deviation of the amplitude ratio $A/A^\prime$ from unity) and the requirement that the temperature be an analytic function of $z/L$ throughout the sample.

\subsubsection{Radial dependence of $\Theta$}

\begin{figure}
\begin{center}
\includegraphics[width=0.8\textwidth]{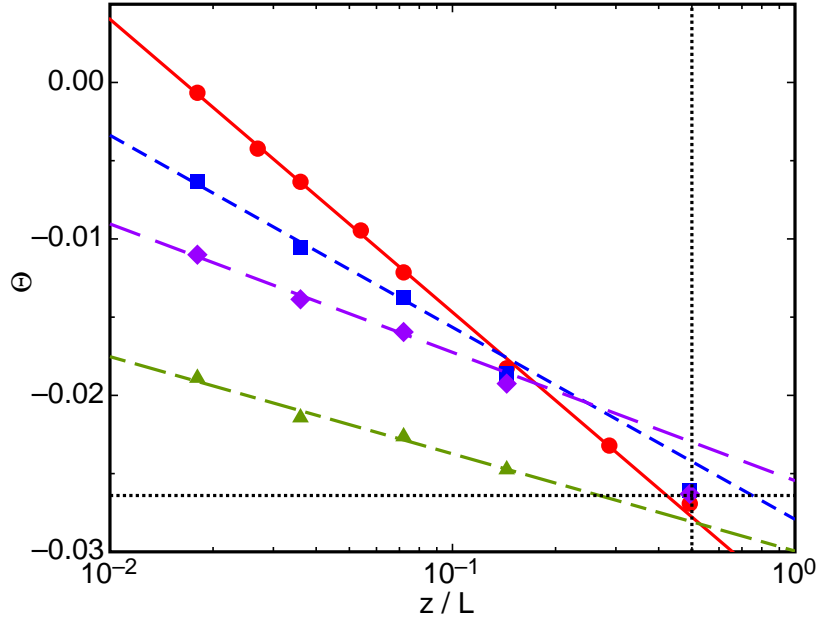}
\caption{Results at the four radial locations for $\Theta(z,r)$ from run 1207281 ($\Ra = 9.94\times 10^{14}, \Pra = 0.858$). Near the left edge of the graph we have, from top to bottom, $\xi = (R-r)/R = 0.0180, 0.0356, 0.0716$, and 0.2680.}
\label{fig:radial}
\end{center}
\end{figure}

\begin{figure}
\begin{center}
\includegraphics[width=0.8\textwidth]{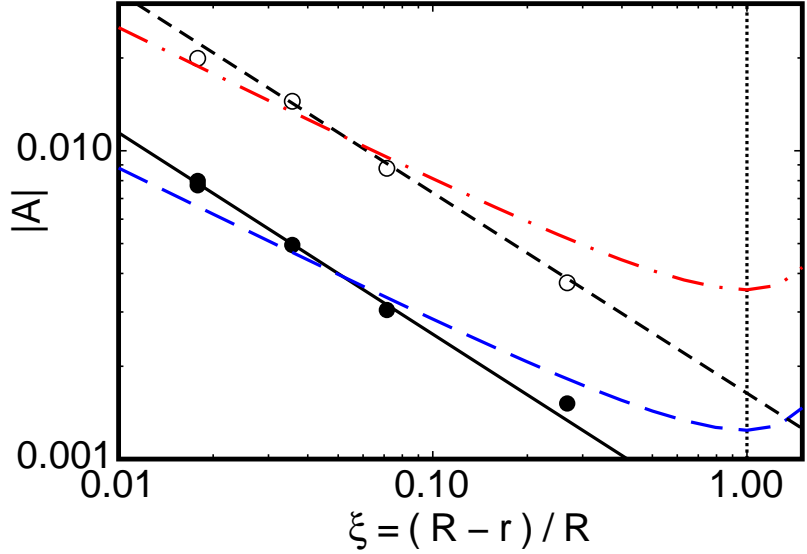}
\caption{The absolute value  $|A|$ of the amplitude  of the logarithmic temperature profiles shown in Figs.~\ref{fig:radial}  and \ref{fig:radial2}. Open circles: $\Ra = 2.0\times 10^{12}$, classical state. Solid circles: $\Ra = 1.049\times 10^{15}$, ultimate state.
Short dashed line: $|A| = 0.0016*\xi^{-0.65}$. 
Solid line: $|A| = 0.00056*\xi^{-0.65}$. 
Dash-dotted line (red online): Eq. \ref{eq:GL}  with $A_1 = 0.00354$.
Long dashed line (blue online): Eq. \ref{eq:GL}  with $A_1 = 0.00124$.
The vertical dotted line is located at the axis of the sample.}
\label{fig:radial3}
\end{center}
\end{figure}

In Fig.~\ref{fig:radial} we show results for $\Theta(z/L)$ at four different radial positions $\xi = (R-r)/R$ for $\Ra = 9.94\times 10^{14}$. As for the classical state illustrated by Fig.~\ref{fig:radial1}, one sees that the amplitude $A$ of the logarithm ({\it i.e.} the slope of the lines fit to the data) decreases as the radial position moves away from the side wall into the sample interior.
Remarkably, all four data sets yield the same value $\phi = \Theta(1/2) = -0.0267\pm 0.0004$. Thus, as in the classical state,  the measurements reveal no measurable deviation from a constant for the temperature in the horizontal mid-plane of the sample.

Figure ~\ref{fig:radial3} shows the absolute value $|A|$ of the amplitude of the logarithmic term (solid circles) as a function of $\xi$ on double logarithmic scales. Also shown for comparison  are the results for the classical state at $\Ra = 2.0\times 10^{12}$ from Fig.~\ref{fig:radial2} (open circles). One sees that both data sets are described quite well by power laws, with very similar exponents. We also show the \cite{GL12} prediction Eq.~\ref{eq:GL} for aspect ratio $\Gamma =1$, with $A_1 = 0.0035$  (dash-dotted line, red online) as shown already in Fig.~\ref{fig:radial2}, and with $A_1 = 0.0012$ (dashed line, blue online) which passes through the ultimate state data. One sees that in both cases the fit is not very good; but as pointed out by \cite{GL12}, in their model the more nearly elliptic path of any large-scale circulation in a sample with $\Gamma = 0 .5$  would tend to produce a more rapid radial variation of $|A|$.

\subsection{Dependence of the parameters on \Ra}

\begin{figure}
\centerline{\includegraphics[width=0.85 \textwidth]{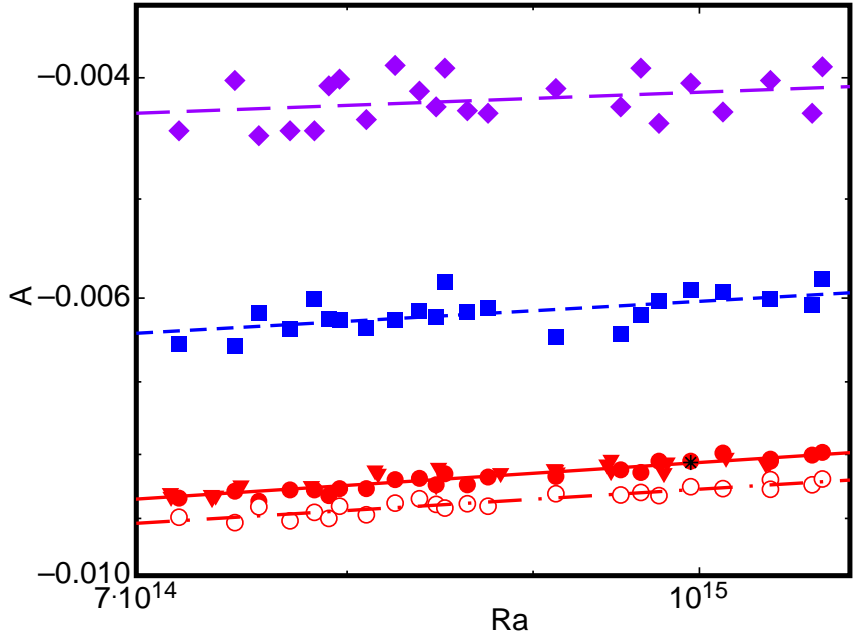}}
\caption{The amplitude $A$ of the logarithmic dependence of $\Theta$ on $z/L$ for the ultimate state as a function of the Rayleigh number \Ra.  Circles and down-pointing triangles (red online): $\xi = 0.0180$. Squares (blue online): $\xi = 0.0356$. Diamonds (purple online): $\xi = 0.0716$. Solid symbols: $A$ from fits to data with $z/L < 0.08$. Open circles: $-A^\prime$ from fits to data with $1-z/L < 0.08$.  The black star corresponds to the data in Fig.~\ref{fig:log_profile}.} 
\label{fig:AofR_ult}
\end{figure}

In the ultimate state it is difficult to establish the \Ra-dependence of any parameter with meaningful accuracy because the amplitudes of the logarithmic term are smaller than in the classical state, and because the data cover at best a factor of two in \Ra. Nonetheless the results for the amplitude $A$ of the logarithmic term given in Fig.~\ref{fig:AofR_ult} indicate that the magnitude of $A$ decreases as \Ra\  increases at all three radial positions shown. This is similar to the \Ra\ dependence found for the classical state.

\begin{figure}
\centerline{\includegraphics[width=0.85 \textwidth]{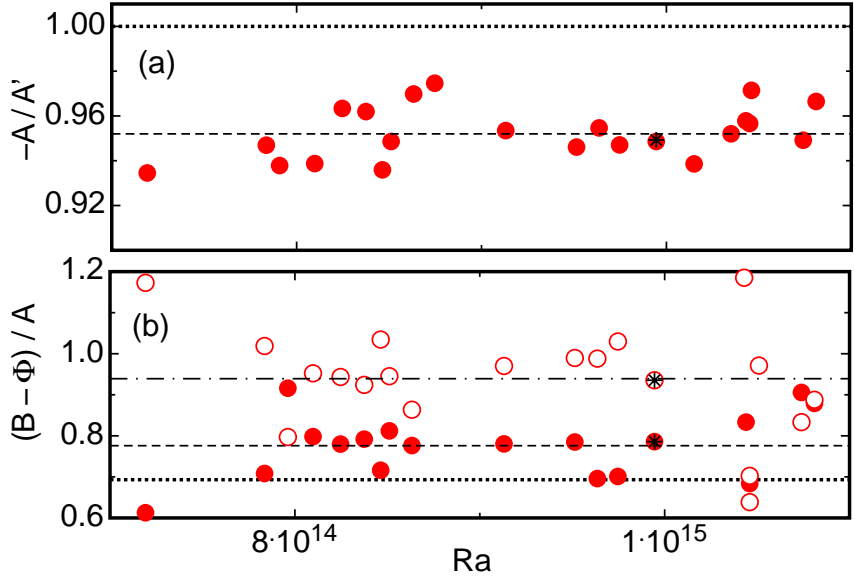}}
\caption{(a): The parameter ratio $-A/A^\prime$ and (b): the ratios $(B-\phi)/A$ (solid circles) and $(B^\prime - \phi)/A^\prime$ (open circles) of the logarithmic dependence of $\Theta$ on $z/L$ for the ultimate state as a function of the Rayleigh number \Ra. In (a) the short dashed line corresponds to the mean value $A/A^\prime = 0.952\pm 0.012$. In (b) the dotted horizontal line correspond to $(B-\phi)/A = - \ln(1/2) \simeq 0.693$, and the short-dashed and dash-dotted lines correspond to the average values $\langle (B-\phi)/A \rangle = 0.776$ and $\langle (B^\prime-\phi)/A^\prime \rangle = 0.94$ respectively. The results for the run shown in Fig.~\ref{fig:log_profile} are indicated by the black stars.}
\label{fig:paras}
\end{figure}

For the radial position $\xi = 0.0180$ an extra effort was made to obtain the best possible statistics by completing two measurements sequences, one starting with run 1207231 (circles) and the other with run 1209171 (triangles). There is no significant difference between the results. A fit of Eq.~\ref{eq:A}  to all those data (solid circles and triangles)  yielded 
\be
A_0 = -4.8\pm 3.0;\ \ \ \  \eta = 0.185\pm 0.018\ . 
\label{eq:A0u}
\ee
A fit of Eq.~\ref{eq:Aprime} to the open circles, for the same $\xi$ but for $(1-z/L) < 0.08$, gave 
\be
A_0^\prime = -3.7\pm 2.1;\ \ \ \  \eta^\prime = 0.176 \pm 0.017\ . 
\label{eq:A0uprime}
\ee
At the other radial positions the scatter is large and it is not possible to determine $\eta$ with good accuracy. The power-law fits yield $\eta = 0.18\pm 0.05$ for $\xi = 0.0356$ (squares) and $\eta = 0.14 \pm 0.13$ for $\xi = 0.0716$ (diamonds). 

We conclude that the above analysis does not reveal any radial dependence of $\eta$, and for $\xi = 0.0180$ (where the accuracy is best) all the data are described well by $\eta = \eta^\prime = 0.18 \pm 0.02$. This exponent value differs by nearly three standard deviations from the result Eqs.~\ref{eq:A0} and \ref{eq:A0prime} for the classical sate; thus we regard it as likely that the exponents of the power-law representations of $A(\Ra)$ differ in the two states. When $\eta=\eta^\prime$ is fixed at 0.18, we find $A_0 = -4.08\pm 0.01$ and $A^\prime_0 = 4.27\pm 0.02$, giving $A_0/A_0^\prime = -0.955\pm 0.006$. Thus the difference between the amplitudes $A$ and $A^\prime$ is significant. This differs from the result $A/A^\prime = -0.988 \pm 0.005$ for the classical state (see sec.~\ref{sec:para_on_Ra}), which did not rigorously rule out  the symmetric case $A = -A^\prime$.

One can also examine the amplitude ratio for individual \Ra\ values. This is done in Fig.~\ref{fig:paras}(a). The data reveal no \Ra\ dependence for $-A/A^\prime$. The average of the data in the figure is 0.952 with a standard deviation of 0.012. This result is consistent with the one cited in the previous paragraph, and both analyses indicates that the OB reflection symmetry about the horizontal mid plane is broken.

In Fig.~\ref{fig:paras}(b) we show $(B^\prime - \phi)/A^\prime$  (open circles) and $(B-\phi)/A$ (solid circles) as a function of \Ra. Although there is considerable scatter, the data do not reveal any dependence on \Ra\  and give the average values $\langle(B^\prime - \phi)/A^\prime\rangle = 0.939$ corresponding to $Z_0^\prime = 0.39$ and $\langle(B-\phi)/A\rangle = 0.776$ corresponding to $Z_0 = 0.46$. Thus we see that the log layer in the lower half of the sample extends essentially to the horizontal mid plane ($Z_0 = 1/2$), while in the upper half it extends less far and space remains for an outer layer. The value $Z_0^\prime = 0.39$ is similar to the result $Z_0^\prime = 0.36\pm 0.01$ found in the classical state (see Sec.~\ref{sec:devlog}). Thus, also this parameter indicates a lack of reflection symmetry about the mid plane for the ultimate state.

\section{Summary}
\label{sec:summary}

In this paper we presented detailed measurements of the logarithmic temperature profiles  for both the classical and the ultimate state of turbulent RBC over the ranges $6\times 10^{11} \alt \Ra \alt 5\times10^{12}$ and $7\times 10^{14} \alt \Ra \alt 1.1\times10^{15}$ respectively.

Even though non-OB effects shifted the center temperature slightly away from the mean temperature, we found for the classical state that the amplitudes $A$ and $A^\prime$ are equal within our resolution. We show that deviations from the logarithmic profile occur for $z/L \agt 0.1$ and $1-z/L \agt 0.1$, indicating the existence of an outer layer beyond the log layer. The widths of the log and the outer layer are (within our resolution) independent of \Ra\ and symmetric about the horizontal mid plane. We discuss the radial dependence of $A$ and $A^\prime$ in terms of a power-law dependence which leads to an effective exponent of about 0.65, and in terms of a functional form proposed by \cite{GL12} for the ultimate state.

For $\Ra = 2\times 10^{12}$ the agreement of DNS with the measurements at the same \Ra\ is remarkably good at radial positions not too near the centerline.  

In Sec.~\ref{sec:para_on_Ra} we presented the dependence of $A$ on \Ra. There we found that  $A(\Ra,\xi = 0.018) = A_0 \Ra^{-\eta}$ with $\eta \simeq 0.12$ (at other $\xi$ the data were not sufficiently reliable to determine $\eta$ with meaningful accuracy). This result differs from the analogous universal inverse von K\'arm\'an constant of shear flow which is independent of the Reynolds number. Although it agrees quite well with a recent prediction by \cite{SCCZBH14}, we note that very recent as yet unpublished measurements for a much larger $\Pra = 12.3$ by P. Wei and G. Ahlers yielded a \Ra-independent $A$ and $A^\prime$. Thus, it seems that the analogy to a universal shear-flow von K\'arm\'an constant is perhaps recovered in the large-\Pra\ limit.

In Sec.~\ref{sec:compLoW} we concluded the discussion of the classical state with an outline of the analogies, similarities, and differences between RBC and shear flows. There we show that the domain diagram for RBC, with a thermal boundary layer, a buffer layer, a log (inner) layer, and an outer layer, is very similar to the equivalent diagram for shear flows when, in the RBC case, length is scaled by the thermal BL thickness $\lambda_{th}$. We argue that $\lambda_{th}$ is the natural inner length scale for near-wall phenomena in RBC because it is presumed to determine the width of the thermal excitations, {\it i.e.} of the plumes which in RBC, we believe, play a role similar to the coherent eddies in shear flow.

In Sec.~\ref{sec:ult} we presented results for the ultimate state in the same sequence as had been done in Sec.~\ref{sec:classical} for the classical state. We shall not describe them in detail here because quiet remarkably, at a qualitative level, the observed phenomena are very similar to those found below $\Ra^*$, and even quantitatively there are few distinctions. However, some important differences do become apparent. Thus, within our resolution, the dependence of $A$ on \Ra\  yields a slightly larger exponent $\eta \simeq 0.18$. This value is significantly larger than the effective-exponent prediction $\eta_{GL} \simeq 0.043$ of \cite{GL12} for the ultimate state but close to the value 0.162 predicted by \cite{SCCZBH14} which, they argue, should apply both in the ultimate and the classical state. We find that the log layer in the lower half of the sample extends up to the horizontal mid plane, as had been predicted by \cite{GL12}. In the upper half the log amplitude $A^\prime$ is slightly smaller than $A$, suggesting an influence of deviations from the OB approximation. Given the unequal log amplitudes, it is not possible for both log layers to meet in the sample center because this would lead to a singularity in the temperature profile; thus, as originally suggested by S. Grossmann (private communication) the required analyticity of $\Theta(z)$  may explain the deviations from the log profile in the upper half when $1-z/L \agt 0.3$ or so. 

\begin{acknowledgments}

\section{Acknowledgment}
We thank Andreas Kopp, Artur Kubitzek, and Andreas Renner for their enthusiastic technical support. We are very grateful to Holger Nobach for many useful discussions and for his contributions to the assembly of the experiment. We benefitted from numerous discussions with S. Grossmann and D.Lohse. We are very grateful to the Max-Planck-Society and the Volkswagen Stiftung, whose generous support made the establishment of the facility and the experiments possible. We thank the Deutsche Forschungsgemeinschaft (DFG) for financial support through SFB963: ``Astrophysical Flow Instabilities and Turbulence".  The work of G.A. was supported in part by the U.S National Science Foundation through Grant DMR11-58514. 
\end{acknowledgments}

\vfill\eject

%\bibliographystyle{jfm}
%\bibliography{./refs}

 \end{document}